\documentclass[showpacs,twocolumn,aps]{revtex4}
\usepackage{amssymb}
\usepackage{amsmath}
\usepackage{graphicx}
\usepackage{lscape}
\usepackage{booktabs}
\usepackage[misc]{ifsym}
\begin{document}
\title{Light (anti-)nuclei and (anti-)hypertriton production\\in $pp$ collisions at $\sqrt{s} =0.90, 2.76$ and $7$~TeV }

\author{\footnotesize Nserdin A. Ragab$^{1,2}$, Zhi-Lei She$^{1,2}$, Gang Chen$^{2,}$\footnote{Corresponding Author: chengang1@cug.edu.cn}}
\affiliation{%
$^{1}$Institute of Geophysics and Geomatics, China University of Geosciences, Wuhan 430074,China\\
$^{2}$School of Mathematics and Physics, China University of Geosciences, Wuhan 430074, China}

\begin{abstract}
Production of light (anti-)nuclei and (anti-)hypertriton within midrapidity ($|y|<0.5$) and $p_T<3.0$~GeV/c in $pp$ interactions at $\sqrt{s}$ = 0.90, 2.76 and 7 TeV is investigated by the dynamically constrained phase space coalescence model (DCPC), combined with {\footnotesize{PACIAE}} model. The ALICE data for yields, ratios, as well as transverse momentum distributions of $\overline{d}$ and $d$ are well reproduced by the model simulations, meanwhile the three basic characters of $\rm^3{\overline{He}}$, $\rm^3{{He}}$, $\rm{{^3_{\overline \Lambda}\overline H}}$, and $\rm{^3_\Lambda H}$ are also predicted. Besides, we found the yields of light (anti-)nuclei produced are dependent upon their mass number $A$, namely, their yields sharply decrease with the increasing of $A$. The strangeness population factor $s_3=\rm{({^3_{\Lambda}H}/{^3{He}})/(\Lambda/p)}$ is found to be about 0.7 $\sim$ 0.8, and is comparable with the available experimental results.
\end{abstract}

\maketitle

\section{Introduction}

The investigation of anti-nuclei is a hot and frontier topic in particle and nuclear physics, cosmology, and astrophysics. Since plentiful nuclei and anti-nuclei can be produced in high energy accelerator experiment, it exactly provides a chance to study light (anti-)nuclei production~\cite{Keane}.

In the last decades, production of antimatter has been a focus in several collision systems under different energies. In large systems, the light anti-nuclei (e.g., $\overline{d}$, $\rm^3{\overline{He}}$, even $\rm^4{\overline{He}}$) and anti-hypertrition ($\rm{{^3_{\overline \Lambda}\overline H}}$) have been successfully measured, such as in Au-Au collisions~\cite{PHE1,star5,star3,star2} within $\sqrt{s_{NN}} = 7.7$~GeV to 200~GeV and in Pb-Pb interactions~\cite{ALI2,Adam,ALI6,53} at $\sqrt{s_{NN}} = 2.76$~TeV, respectively. For the little systems, ALICE Collaboration has also published papers on light anti-nuclei production in $pp$ interactions ~\cite{ALI2,Adam,ALI3} at $\sqrt{s}= 0.90, 2.76$, and $7$~TeV.

Theoretically, one can normally select several numerical models(e.g., a transport model) to predict the production of (anti-)nucleons and (anti-)hyperons. Then, light (anti-)nuclei and (anti-)hypernuclei production are computed with a statistical method~\cite{LiLu2} or a certain phase-space coalescence model~\cite{LiLu1,CLW1,Ma1,Ma2}. As examples some researches applied the selected coalescence model to combine a multiphase transport ({\footnotesize{AMPT}}) model~\cite{Ko1} or the blast-wave approach~\cite{Ma3,Ma4,Ma5}, to study light (anti-)nuclei production in high energy nuclear-nuclear collisions. Besides those efforts, the {\footnotesize{DCPC}} model~\cite{Yan1,chen1} was created to investigate production of light (anti-)nuclei~\cite{chen2,chen3} in high energy $pp$ collisions~\cite{chen4} and  nucleus-nucleus (Pb-Pb~\cite{chen5}, and Au-Au~\cite{chen6}, Cu-Cu~\cite{chen7}) collisions, in which the final state hadrons produced within parton and hadron cascade model ({\footnotesize{PACIAE}})~\cite{sa1}.

In this work, we generate final-state hadrons produced by the {\footnotesize{PACIAE}} model in high energy $pp$ collisions(non-single diffractive(NSD) process) at three separate center-of-mass (c.m.) energies. Next, the {\footnotesize{DCPC}} model was applied to study production of light nuclei ($d$, $\rm^3He$ and $\rm^3_\Lambda H$) and their corresponding anti-nuclei ($\overline d$, $\rm^3{\overline {He}}$, and $\rm{{^3_{\overline \Lambda}\overline H}}$). In Sec. 2, we have introduced the {\footnotesize{PACIAE}} model and {\footnotesize{DCPC}} model briefly. In Sec. 3, the model simulations are shown. The Sec. 4 presents a concise plot summary.

\section{Models}

Based on the {\footnotesize {PYTHIA6.4}} model~\cite{Sjostrand}, the {\footnotesize{PACIAE}}~\cite{sa1} is promoted to mainly study the hadron-hadron and nucleus-nucleus collisions, relying on the collision geometry and nucleon-nucleon ($NN$) total cross section. In the {\footnotesize{PACIAE}} model the strings at this stage will randomly break into free partons, forming the partonic initial state.
Then parton rescattering is introduced using the 2 $\rightarrow$ 2 (LO-pQCD) interaction cross sections of parton-parton ~\cite{Cambridge}.
After parton rescattering the hadronization then proceeds~\cite{Yan, Sjostrand,sa1}.
At last, a hadron rescattering is introduced, in which the two-body collision method~\cite{Tai,Sa} is applied, until all hadrons have reached freeze-out.

We generated the final-state particles by {\footnotesize{PACIAE}} model~\cite{sa1}, and then hadrons are processed with {\footnotesize{DCPC}} model~\cite{Yan1} to build light (anti-)nuclei. Considering the uncertainty principle in quantum statistical mechanics~\cite{kubo},
\begin{equation}
\Delta\vec q\Delta\vec p\geq h^3,
\end{equation}
 one can only know that a particle (positions $\vec q$ and momentum $\vec p$ ) lies somewhere within a volume of $\Delta\vec q\Delta\vec p$ or state inside a quantum "box" occupied a volume ($h^3$) of six-dimension phase space~\cite{chen1}. Then one could define a integral to directly evaluate the yield for a single particle:
\begin{equation}
Y_1=\int_{H\leq E} \frac{d\vec qd\vec p}{h^3}.
\end{equation}
Here, $H$ is the Hamiltonian, $E$ denotes the energy of the particle. Likewise, the yield for N particles could be valued with
\begin{equation}
Y_N=\int ...\int_{H\leq E} \frac{d\vec q_1d\vec p_1...d\vec q_Nd\vec
p_N}{(h^3)^{N}}. \label{phas}
\end{equation}
While Eq.(3) should satisfy the following constraint conditions:
 \begin{equation}
  \begin{array}{ll}
    \hspace{0.2cm} |\vec q_{ij}| \leqslant D_0, (i\neq j;i,j = 1,2,3,...,N),\\
    \hspace{0.2cm} m_0-\Delta m \leqslant m_{inv}\leqslant m_0+\Delta m,

  \end{array}
\end{equation}
where
\begin{equation}
m_{inv}=\sqrt{(E_1+E_2+E_3)^2-(\vec p_1+\vec p_2+\vec p_3)^2},
\label{yield2}
\end{equation}
$|\vec q_{ij}|$ denotes the distance between particles $j$-th and $i$-th, $D_0$ and $m_0$ are the diameter and rest mass of N particles cluster, respectively. $p_i$ and $E_i$ ($i = 1, 2, 3$) represent the momenta and energies of particles, respectively. $\Delta m$ stands for the allowed uncertainty. In Eq.(3), the integral of continuous distributions will be changed by using the sum for discrete distributions, because the hadron momentum and position distributions are discrete in transport model simulation.

\section{Results and Discussion}

We generate the final-state hadrons by the transport model {\footnotesize{PACIAE}}. Then utilize the {\footnotesize{DCPC}} model to coalescence production of light (anti-)nuclei~\cite{Yan1}. The model parameters are fixed on the default values given in {\footnotesize{PYTHIA6.4}}. However, the $K$ factor
as well as the parameters parj(1), parj(2), and parj(3) were roughly fitted the (anti-)proton and $\Lambda$($\bar{\Lambda}$) from ALICE data in mid-rapidity $pp$ collisions at $\sqrt{s}= 0.90, 2.76$ and $7$ TeV~\cite{Sjostrand,B3,Aamodt,Adam,jadam}, as shown in Fig. 1. It shows that the results of {\footnotesize{PACIAE}} simulation are very close to the  ALICE data. The parameters were chosen from figure 1 as parj(1) = 0.07 (default value is 0.10), parj(2) = 0.18 (0.30), parj(3) = 0.40 (0.40), with $K = 0.95$ (1.0 or 1.5), which were applied to compute the yields of $d, \overline d$, $\rm^3{He}, \rm^3{\overline {He}}$, $\rm{^3_{\Lambda}H}$, and $\rm{{^3_{\overline \Lambda}\overline H}}$, in mid-rapidity $pp$ collisions of the c.m energy of 0.90, 2.76 and 7 TeV relying on the final-state hadrons from the {\footnotesize{PACIAE}} simulations. The consequence of light (anti-)nuclei yields are shown on Table 1.

\begin{figure}[th]
\centerline{\includegraphics[width=4.2in]{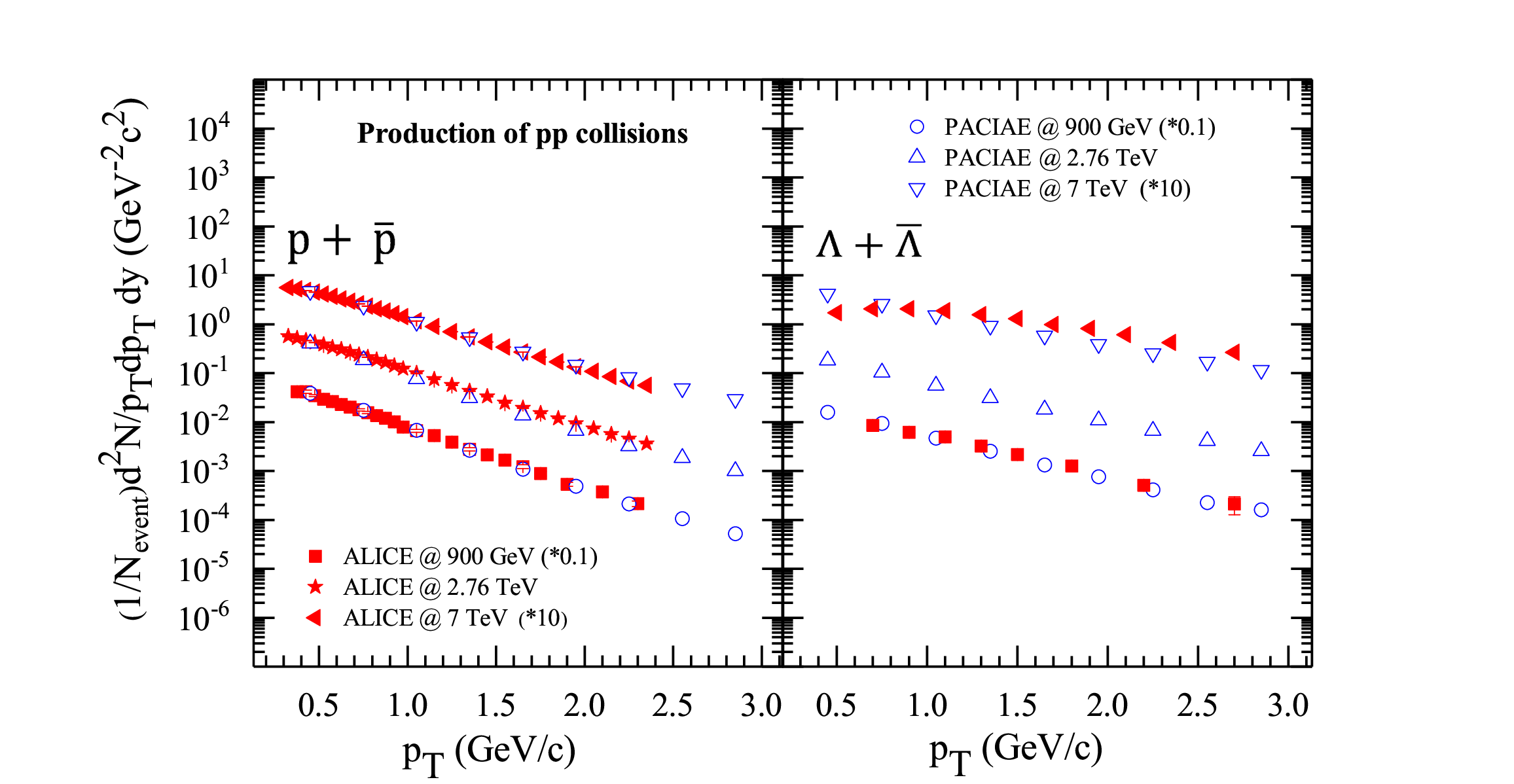}}
\vspace*{8pt}
\caption{The transverse momentum ($p_T$) distribution of $p + \bar{p}$ and $\Lambda + \bar{\Lambda}$ in mid-rapidity $pp$ collisions at c.m. energy of 0.90, 2.76 and 7 TeV, calculated by {\footnotesize{PACIAE}} model. The data have been multiplied by constant factors inside the figure. The ALICE data are taken from~\cite{Aamodt,B3,Adam,jadam}.\label{f1}}
\end{figure}

 One can see in the Tab.1 that:
\begin{itemize}
\item The yield computed by {\footnotesize{PACIAE}} model of $\rm{^3_{\Lambda}H}$($\rm{{^3_{\overline \Lambda}\overline H}})$ is significantly less than that of $\rm{^3He}(\rm^3{\overline{He}})$, since the yield of hyperons is less than that of protons.
\item When the c.m. energy varies from 900 GeV to 7 TeV, the (anti-)proton yield calculated by {\footnotesize{PACIAE}} simulations increases $\sim$ 50\%. That is less than the increase of light (anti-)nuclei (over 100\% for $d$ and $ \overline d$, $\sim$ 60\% for $\rm^3{He}$ and $\rm^3{\overline {He}}$). Here, the yields of $d$($pn$) and $\rm^3{He}$($ppn$) produced by nucleon combination will inevitably increase when the c. m. energy of $pp$ collision increases and the number of final nucleons increases. However, this may attribute to a larger increase of available phase space for anti-nuclei production than that for anti-hadron production.
\item The yield of $d$ and $\bar{d}$ calculated by {\footnotesize{PACIAE+DCPC}} simulation nicely matches the ALICE data in the range of uncertainty. Meanwhile, we predict the yield of $\rm{^3He}$, $\rm^3{\overline {He}}$, $\rm{^3_{\Lambda}H}$, and $\rm{{^3_{\overline \Lambda}\overline H}}$ in rapidity $pp$ collisions at c.m. energy of 0.90, 2.76 and 7 TeV using {\footnotesize{PACIAE+DCPC}} model.
\end{itemize}

 It should be noted that the yield of $d$ and $ \overline d$ is simulated with the parameter $\Delta m = 0.003$~GeV, and the yield of $\rm{^3_{\Lambda}H}$($\rm{{^3_{\overline \Lambda}\overline H}})$ and $\rm{^3He}(\rm^3{\overline{He}})$ is simulated with the parameter $\Delta m = 0.0075$~GeV. The parameter $\Delta m$ for $d(\overline d)$ and $\rm{^3He}(\rm^3{\overline{He}}, \rm{^3_{\Lambda}H},\rm{{^3_{\overline \Lambda}\overline H}})$ are determined by fitting the yield of $d$ and $\rm{^3He}$ with ALICE experimental data in $pp$ collisions at c.m. energy of 7~TeV~\cite{ALI3}, respectively.

\begin{large}
\small\addtolength{\tabcolsep}{7.pt}
\begin{table*}[t]
\caption{The yields of (anti-)particles in mid-rapidity $pp$ collisions at c.m. energy of 0.90, 2.76 and 7 TeV calculated by  {\footnotesize{PACIAE+DCPC}} simulations, compared with the ALICE data~\cite{ALI3,Aamodt,B3,Adam,Abbas,jadam}}.
\begin{tabular}{cccccccc}
\hline 
 \scriptsize Particle &&\scriptsize PACIAE+DCPC  &&&\scriptsize ALICE& \\ \hline
  &0.90 TeV &2.76 TeV    &7 TeV            &0.90 TeV               &2.76 TeV              &7 TeV \\ \hline
$p$                       &0.082	&0.090	    &0.124            &$0.083\pm0.008$	     &$0.090\pm0.007$	    &$0.124 \pm 0.009$\\
 $\bar{p}$                &0.079	&0.088	    &0.122            &$0.079\pm0.008$	     &$0.088\pm0.006$	    &$0.123\pm0.010$\\
 $\Lambda$                &0.048	&0.060	    &0.087            &$0.048\pm0.005$	      &--	                 &$0.090\pm0.007$  \\
  $\bar{\Lambda}$          &0.047	&0.060	    &0.086            &$0.047\pm0.007$	      &--	                 &$0.089\pm0.006$   \\
 $d$$^a$                 &1.06E-4	&1.41E-4	&2.04E-4 	      &$(1.12\pm0.13)$E-4    &$(1.53\pm0.14)$E-4      &$(2.02\pm0.17)$E-4\\
 $\bar{d}$$^a$           &9.83E-5	&1.35E-4	&1.98E-4          &$(1.11\pm0.13)$E-4    &$(1.37\pm0.13)$E-4      &$(1.92\pm0.15)$E-4\\
 $\rm{^3He}$$^b$            &5.17E-8	&7.79E-8	&1.16E-7           &--	                  &--	                &--\\
 $\rm^3{\overline {He}}$$^b$ &4.62E-8	&7.28E-8	&1.10E-7           &--	                  &--	                &$(1.10\pm0.63)$E-7\\
 $\rm{_{\Lambda}^3 H}^c$      &2.36E-08	           &3.69E-08	      &5.62E-08          &--	        &--                   &--  \\
 $\rm{{^3_{\overline \Lambda}\overline H}}^c$ &2.06E-08	&3.32E-08	       &5.23E-08          &--	        &--      &-- \\ \hline
\end{tabular}
\label{paci1}
\end{table*}
\end{large}

\begin{figure*}[htb]
\includegraphics[width=0.8\textwidth]{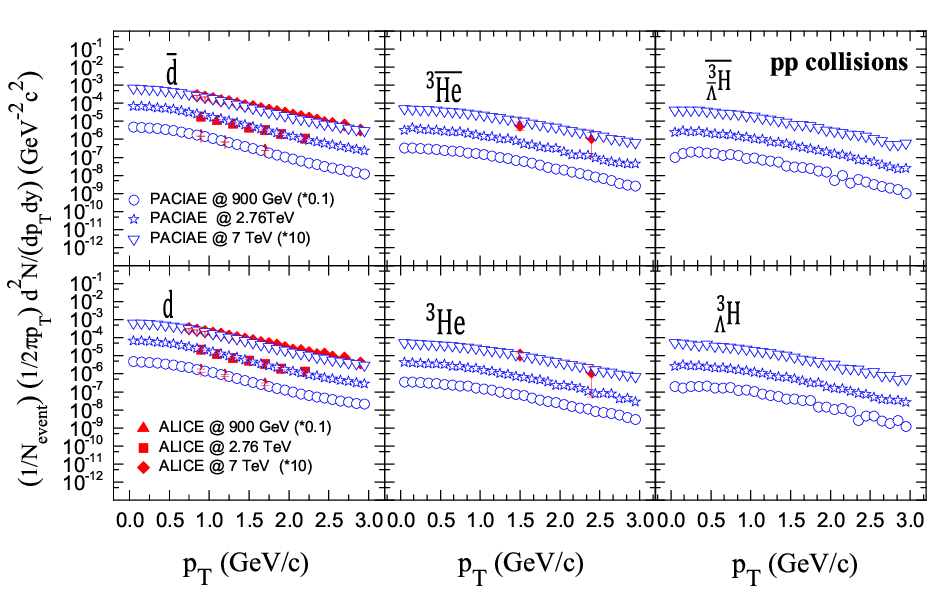}
\caption{The transverse momentum distribution of $d,\overline{d}$ ,$\rm{^3He}$, $\rm{\overline{^3He}}$, $\rm{^3_{\Lambda}H}$, and $\rm{{^3_{\overline \Lambda}\overline H}}$ computed by {\footnotesize{PACIAE+DCPC}} simulations in rapidity $pp$ collisions at c.m. energy of 0.90, 2.76 and 7~TeV. The solid points are ALICE data~\cite{ALI3}; for clarity, the data are separated with multiplying constant factors here.\label{f2}}
\end{figure*} 

Fig. 2 presents the spectrum of integral yields for $d(\bar{d})$, $\rm{^3He}(\rm^3{\overline{He}})$, and $\rm{^3_{\Lambda}H}$($\rm{{^3_{\overline \Lambda}\overline H}}$) calculated by {\footnotesize{PACIAE+DCPC}} model with $ p_T < 3$~GeV/c, $|y|<0.5$ in $pp$ collisions at $\sqrt{s}= 0.90, 2.76 $ and $7$~TeV, individually. One can see the transverse momentum spectrum of $d, \bar{d}$, $\rm^3He$, $\rm^3{\overline {He}}$ calculated by {\footnotesize{PACIAE+DCPC}} simulations is consistent with the result distribution of {\footnotesize{ALICE}} data~\cite{ALI3}. Then we predict the whole distribution of the transverse momentum spectrum of $\rm{^3He}$, $\rm^3{\overline {He}}$, $\rm{^3_{\Lambda}H}$, and $\rm{{^3_{\overline\Lambda}\overline H}}$  in $pp$ collisions at these three separate energies, respectively.

\begin{table*}[th]
\small\addtolength{\tabcolsep}{7.pt}
\caption{The top part show the ratio of anti-particles ($\bar{p}, \bar{\Lambda}$, $\bar{d},\rm^3{\overline {He}}$, and $\rm{{^3_{\overline \Lambda}\overline H}}$) to particles ($p, \Lambda, d, \rm{^3He}$, and $\rm{{^3_\Lambda H}}$) in rapidity $pp$ collisions at $\sqrt{s}= 0.90, 2.76 $ and $7$~TeV, then display the mixed ratios between different (anti-)particles. On the bottom the ratios of $\Lambda$($\bar{\Lambda}$) to $p(\bar{p})$ and $\rm{^3_{\Lambda}H}$($\rm{{^3_{\overline \Lambda}\overline H}}$)  to $\rm{^3He}(\rm^3{\overline{He}})$ are presented. ALICE data are from Ref.~\cite{ALI3,Abbas}.}
\begin{tabular}{cccccccccc} \hline
 Particle &&{\scriptsize PACIAE+DCPC }&&&{\scriptsize ALICE}&\\ \hline
 &0.90 TeV &2.76 TeV&7 TeV  &0.90 TeV &2.76 TeV &7 TeV \\ \hline
 $\bar{p}$/$p$ &0.963	&0.978	&0.984	&$0.952\pm0.002$ &$0.978\pm0.002$ &$0.992\pm0.009$\\
 $\bar{\Lambda}$/$\Lambda$ &0.981	&0.993	&0.984	&$0.963\pm0.023$ &$0.979\pm0.015$ &$0.989\pm0.014$\\
 $\bar{d}$/$d$  &0.927	&0.954	&0.967 &$0.991\pm0.09$	 &$0.895\pm0.05$	 &$0.950\pm0.02$\\
 $\rm^3{\overline {He}}$/$\rm{^3He}$  &0.893	&0.935	&0.948 &--	  &--	     &--\\
 $\rm{{^3_{\overline \Lambda}\overline H}}/{^3_{\Lambda}H}$ &0.873	&0.909	&0.925 &--	  &--	     &--\\ \hline
 $d$/$p$              &1.29E-3 &1.57E-3	&1.65E-3  &$(1.38\pm0.186)$E-3 &$(1.48\pm0.167)$E-3 &$(1.63\pm0.170)$E-3\\
 $\bar{d}$/$\bar{p}$  &1.24E-3 &1.53E-3	&1.62E-3  &$(1.39\pm0.205)$E-3 &$(1.31\pm0.145)$E-3 &$(1.56\pm0.170)$E-3\\
 $\rm{^3He}$/$d$         &4.88E-4 &5.52E-4	&5.69E-4  &--	  &--	     &--\\
 $\rm^3{\overline {He}}$/$\bar{d}$ &4.70E-4 &5.39E-4	&5.56E-4   &--	  &--	     &$(5.73\pm3.26)$E-4\\ \hline
 $\Lambda$/$p$  &0.585 & 0.667	&0.702   &$0.578\pm0.082$  &--	     &$0.726\pm0.077$\\
 $\bar{\Lambda}$/$\bar{p}$ &0.595 & 0.670	&0.705   &$0.595\pm0.107$  &--	     &$0.724\pm0.076$ \\
 $\rm{^3_{\Lambda}H}$/$\rm{^3He}$ &0.456 & 0.474	&0.484   &--	  &--	     &--\\
 $\rm{{^3_{\overline \Lambda}\overline H}}$/$\rm^3{\overline {He}}$ &0.446 & 0.456 &0.475   &--	  &--	     &--\\ \hline
\end{tabular}
\label{paci2}
\end{table*}
\begin{figure*}[th]
\centerline{\includegraphics[width=4.50in]{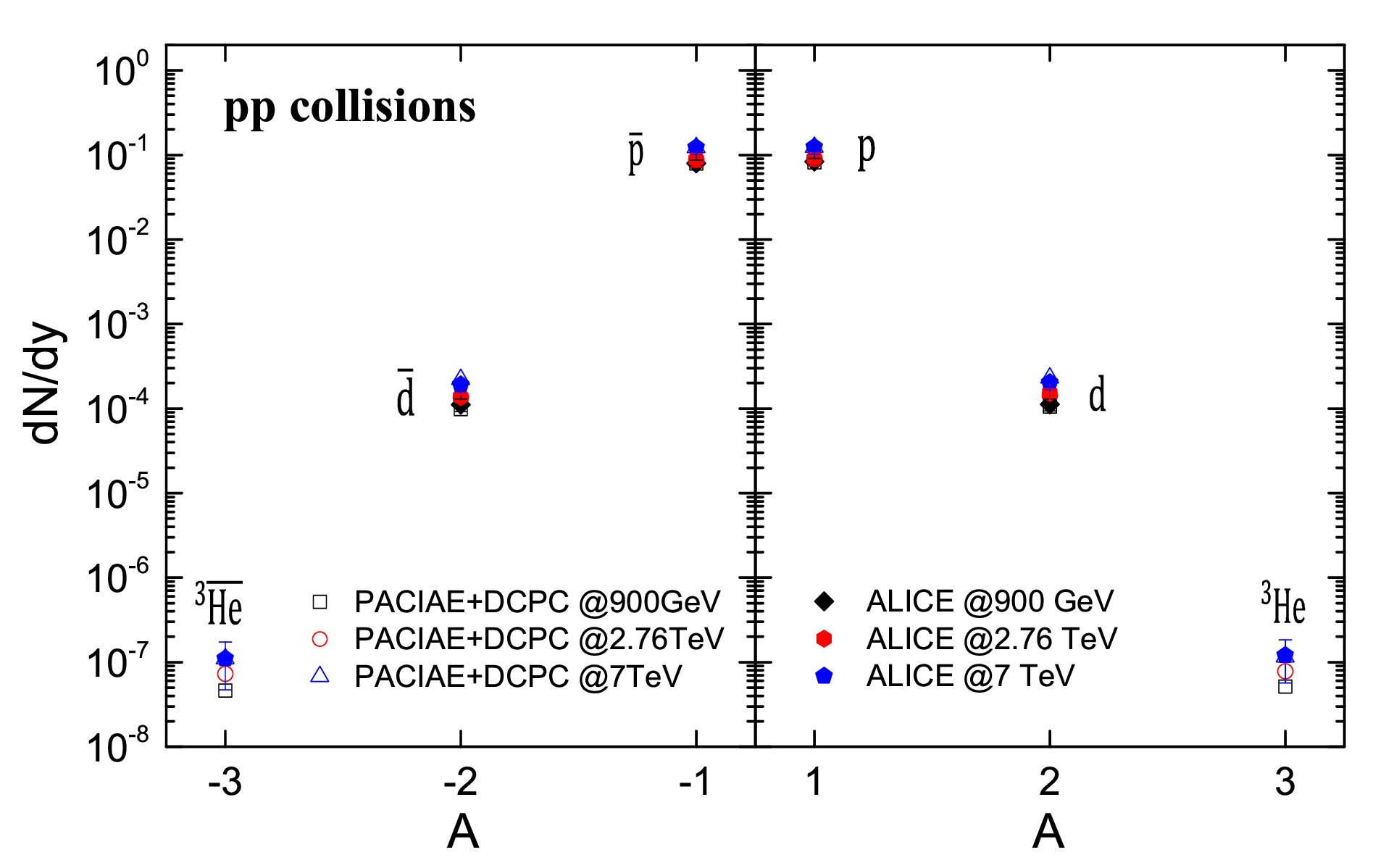} }
\vspace*{-8pt}
\caption{Integrated yields of anti-nuclei (left section) and nuclei (right section) in mid-rapidity $pp$ collisions at $\sqrt{s}$ = 0.90, 2.76 and 7 TeV, vary with atomic mass number $A$. And {\footnotesize{PACIAE+DCPC}} model (see open point) results are displayed. The ALICE data (solid point) is taken from~\cite{ALI3,B3,Aamodt,Adam,jadam,Abbas}.\label{f3}}
\end{figure*}

 Considering high energy collisions, the nature of light (anti-)nuclei production mechanism is final-state hadron coalescence, therefore the yield ratios of nuclei can be predicted through the coalescence model. In theory, the yield ratios for two (anti-)nuclei could be straightly compared with yield ratios of constituent hadrons in the naive coalescence framework~\cite{cley,Ma3}. E.g., the ratio of $\rm{^3_{\overline \Lambda}\overline H}/\rm{^3_{\Lambda}H}$ is supposed to proportional to (${\overline\Lambda}/{\Lambda}$)($\rm{\overline p}/{p}$)($\rm {\overline n}/{n}$), due to  $\rm^3_{\overline \Lambda}\overline H$ and $\rm^3_{\Lambda}H$
are individually generated by ($\rm \overline p+\overline n+\overline\Lambda$) and ($\rm p+n+\Lambda$). Hence one can write the yield ratios as follows:
\begin{equation}
\frac{\rm {^3_{\overline \Lambda}\overline H}}{\rm{^3_{\Lambda}H}}=\\
 \frac{\rm {\overline p \overline n \overline\Lambda}}{\rm{pn\Lambda}}\simeq\\
 (\frac{\rm \overline p}{\rm p})^2\frac{\rm \overline\Lambda}{\Lambda},
\end{equation}
once again, mixed ratios like that:
\begin{equation}
\frac{\rm {^3_{\Lambda}H}}{\rm{^3He}}=\\
 \frac{\rm {pn\Lambda}}{\rm{ppn}}\simeq\\
 \frac{\Lambda}{\rm p}.
\end{equation}

The yield ratio of anti-particles ($\bar{p}$, $\bar{\Lambda}$, $\bar{d}$, $\rm^3{\overline {He}}$ and $\rm{{^3_{\overline \Lambda}\overline H}}$) to their corresponding particles ($p$, ${\Lambda}$, $d$, $\rm{^3He}$, and $\rm{^3_{\Lambda}H}$) and their mixed ratios ($ \rm {\Lambda}/p$, $\rm{\overline \Lambda}/\overline p$, $\rm{{^3_{\Lambda}H}/\rm{^3{He}}}$, $\rm{{^3_{\overline \Lambda}\overline H}/\rm^3{\overline {He}}}$)
in $pp$ collisions at $\sqrt{s}= 0.90, 2.76$ and $7$ TeV are presented in Table II.
Obviously, we can obtain the ratios of anti-nuclei and anti-hypertriton to nuclei and hypertriton are all dependent on the c.m. energy, as same as their yields increase with the c.m. energy increase as the Table I shown. One also find the ratios of anti-particles to their corresponding particles, and (anti-)hypertriton to (anti-)helium-3 ($\rm{{^3_{\Lambda}H}/{^3{He}}}$, $\rm{{^3_{\overline \Lambda}\overline H}/{^3{\overline{He}}}}$) are less than 1, and the latter implies yield of $\rm{^3_{\Lambda}H}$($\rm{{^3_{\overline \Lambda}\overline H}}$) is less than that of $\rm{^3He}(\rm^3{\overline{He}})$.

As the above Eq.(6) shows, since the constituent hadrons of  $\rm{{^3_{\overline \Lambda}\overline H}}$ and $\rm{{^3_{\Lambda}H}}$ are $(\overline {\Lambda} +\rm\overline p + \rm\overline n)$ and $(\Lambda + \rm p + \rm n)$ in coalescence model, the yield ratio of $\rm{{^3_{\overline \Lambda}\overline H}}$/$\rm{{^3_{\Lambda}H}}$  is proportional to $(\frac{\rm \overline p}{\rm p})^2\frac{\rm \overline\Lambda}{\Lambda}$. And then the ratios $\bar{d}$ to $d$ are consistent with $(\bar{p}/p)^2$, the ratios $\rm^3{\overline {He}}/{^3He}$ is approximately the same as $(\bar{p}/p)^3 $, and the ratios $\rm{{^3_{\overline \Lambda}\overline H}}/{^3_{\Lambda}H}$ is just approximately $(\bar{p}^2\bar{\Lambda})/(p^2\Lambda)$ as the Table II shows. The calculated $\rm{{^3_{\overline \Lambda}\overline H}}$/ $\rm{{^3_{\Lambda}H}}$ values well matched the theoretical assumption that $\rm{{^3_{\overline \Lambda}\overline H}}$ and $\rm{{^3_{\Lambda}H}}$ are individually generated with hadrons coalescence of $(\overline {\Lambda} +\rm \overline p + \rm\overline n)$ and $(\Lambda + \rm p + \rm n)$.
Furthermore, the simulation consequences ($\bar{d}/d$) in our model are found to be comparable with the experimental data from ALICE Collaboration in mid-rapidity $pp$ collisions at $\sqrt{s}= 0.90, 2.76$ and $7$ TeV, respectively.

To further understand the mass and/or c.m. energy dependence of production of light (anti-)nuclei, Fig.3 gives integrated yield distributions of light anti-nuclei ($\bar{p}, \bar{d}, \rm^3{\overline {He}}$) and nuclei ($p, d,\rm^3He$), versus the atomic mass number $A (A = 1$ to $3)$ in different c.m. energy of 0.90, 2.76 and 7~TeV, respectively. The hollow points denote our results computed using {\footnotesize{PACIAE+DCPC}} simulation in  mid-rapidity $pp$ collisions at LHC energies, with $p_T < 3.0$~GeV/c and $|y|< 0.5$. The solid points show the data of ALICE experiment~\cite{ALI3,Aamodt,B3,Adam,Abbas,jadam}. We can find from Fig.3 yields of light (anti-)nuclei decrease sharply as atomic mass number $A$ increase. Numerically, integrated yield of light (anti-)nuclei spans about 3-orders of magnitude with noteworthy exponential type. Besides, Fig.3 shows that {\footnotesize{PACIAE+DCPC}} simulations are consistent with the data from ALICE within uncertainties.

\begin{figure}[th]
{\includegraphics[width=3.3in]{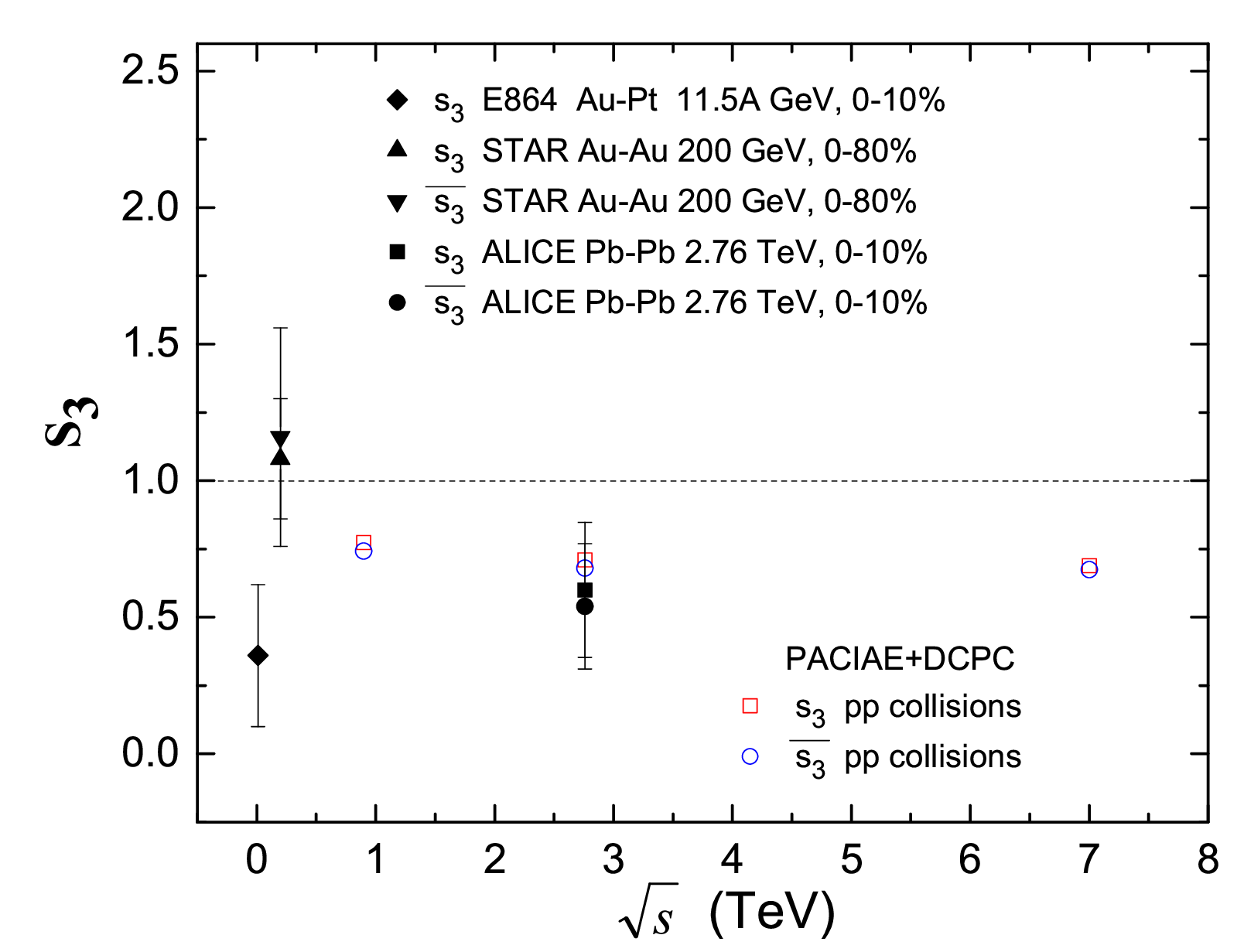}}
\caption{The $s_3$(${\overline s_3}$) value as a function of $\sqrt{s}$ in $pp$ interactions computed by {\footnotesize{PACIAE+DCPC}} model. Here the data from STAR~\cite{star2}, ALICE~\cite{53}, and E864~\cite{52} are also presented, and error bars denote statistical uncertainties.}
\end{figure}

In order to better compare (anti-)hypernuclei with (anti-)nuclei, by analogy with heavy-ion collisions, the strangeness population factor~\cite{52,53} in $pp$ collisions can also be introduced as follows:
\begin{equation}
s_3 = (\rm_{\Lambda}^3H\times p)/(\rm^3{He}\times \Lambda),
\end{equation}
\begin{equation}
\overline{s_3} = (\rm{{^3_{\overline \Lambda}\overline H}}\times \overline p)/(\rm^3{\overline {He}}\times \overline\Lambda).
\end{equation}
Note that, the proton and anti-proton here does not include the contribution of $\Lambda$($\bar{\Lambda}$) decay to proton and anti-proton. The $s_3$ value is likely to study production mechanism for light nuclei in high energy collisions~\cite{Ma1,sato}, due that it's quite related to local baryon-strangeness correlation~\cite{koch,mcheng,braun1}.
The $s_3(\overline s_3)$ values in $pp$ collisions computed by {\footnotesize{PACIAE+DCPC}} model for three separate c.m. energies of 0.90, 2.76 and 7 TeV are presented in Fig.4. The present results for $pp$ collisions show the values of $s_3(\overline s_3)$ are about 0.7 $\sim$ 0.8. Furthermore, the calculated $s_3(\overline s_3)$ values are compared with the results from STAR~\cite{star2}, ALICE~\cite{53}, and E864~\cite{52}, and are consistent with experiment data within uncertainties allowed.

It should be noted that the statistical error of the predicted results calculated by the model is not written, since it is very small and can be ignored.

\section{Conclusion}

The production of light (anti-)nuclei and (anti-)hypertriton has been studied by the {\footnotesize{DCPC}} model, relied on the final-state hadrons created by the {\footnotesize{PACIAE}} model within $|y|<0.5$ and $p_T<3$~GeV/c in mid-rapidity $pp$ collisions at c.m. energy of 0.90, 2.76 and 7 TeV, respectively. The basic variable simulation includes yield, yield ratio, transverse momentum spectrum of $d$, $\bar{d}$, $\rm{^3He}$, $\rm^3{\overline {He}}$, $\rm{^3_{\Lambda}H}$ and $\rm{{^3_{\overline \Lambda}\overline H}}$.

The results of our calculations show a significant dependence of the c.m. energy on yield, ratio, and transverse momentum distribution for $d, \bar{d}$, $\rm^3{{He}}$, $\rm^3{\overline{He}}$, $\rm{^3_{\Lambda}H}$ and $\rm{{^3_{\overline \Lambda}\overline H}}$. When the c.m. energy varies from 900 GeV to 7 TeV, yield of light (anti-)nuclei and (anti-)hypertriton computed with {\footnotesize{PACIAE+DCPC}} simulation increases. The ratios of $\overline{d}$ to $d$, $\rm^3{\overline{He}}$ to $\rm^3{{He}}$, and $\rm{{^3_{\overline \Lambda}\overline H}}$ to $\rm{^3_{\Lambda}H}$ all approach unity with c.m. energies increase, indicating nuclei and anti-nuclei species are produced with similar abundance at LHC energies. The ALICE data for yield, ratio, as well as spectrum of $\overline{d}$ and $d$ are well reproduced with {\footnotesize{PACIAE+DCPC}} simulations. We also obtained yields of light (anti-)nuclei decrease sharply when atomic mass number $A$ increases.
The strangeness population factor $s_3=\rm{({^3_{\Lambda}H}/{^3{He}})/(\Lambda/p)}$ for (anti-)hypertrition with (anti-)helium-3 is calculated to be about 0.7 $\sim$ 0.8, which is compatible with the experimental data.

The data used to support the findings of this study are included within the article.
The authors declare that they have no conflicts of interest.

\section {Acknowledgments}
The work is supported by the NSFC of China under Grant No. 11475149. This work is also supported by the
high-performance computing platform of China University of Geosciences.


\end{document}